# MAI-SIM: interferometric multicolor structured illumination microscopy for everybody

Edward N. Ward, Lisa Hecker, Charles N. Christensen, Jacob R Lamb, Meng Lu, Luca Mascheroni, Chyi Wei Chung, Anna Wang, Christopher J. Rowlands, Gabriele S. Kaminski Schierle, Clemens F. Kaminski

**Sub-diffraction resolution, gentle sample illumination, and the possibility to image in multiple colors make Structured Illumination Microscopy (SIM) an imaging technique which is particularly well suited for live cell observations. Here, we present Machine learning Assisted Interferometric-SIM (MAI-SIM), an easy-to-implement method for high speed SIM imaging in multiple colors. The instrument is based on an interferometer design in which illumination patterns are generated, rotated, and stepped in phase through movement of a single galvanometric mirror element. The design is robust, flexible, and the pattern generation process works for all wavelengths. We complement the unique properties of interferometric SIM with a machine learning toolbox that is simple and efficient to use and is superior to existing methods for the reconstruction of super-resolved images recorded by the instrument. The framework permits real-time SIM reconstructions to be performed in multiple colors, providing the user with instant visualization of the super-resolved images. We demonstrate the capability of MAI-SIM on live biological samples and capture super-resolution images in multiple colors simultaneously over large fields of view. Finally, we embrace a fully open design philosophy to bring the advantages of MAI-SIM to as many users as possible and provide full details on system design and software.**

Super-resolution microscopy has enabled the observation of increasingly smaller features in biological specimens, but many challenging problems remain, for example in the imaging of intracellular transport processes and small organelle interactions in live cells. Reasons for this are limitations in acquisition speed, restrictions on fluorescent labels or required buffers, and the need for use of illumination intensities compatible with viewing live samples. Structured Illumination Microscopy (SIM) stands out as a technique for sub-diffraction imaging of live biological specimens as it strikes an optimal balance in resolution and imaging speed at light intensities where phototoxicity levels are tolerable.[1,2] As SIM uses widefield detection and requires only a linear fluorescent response, large fields of view can be imaged at low excitation powers and high speeds.

In SIM, a spatially modulated illumination pattern is projected onto the sample, which mixes with high spatial frequency content in the sample structure to generate beat frequencies. The low frequency beat patterns, or Moiré fringes, contain high resolution sample information, which can be reconstructed into super-resolved images. In most implementations of SIM, a striped excitation pattern with sinusoidal modulation is used, containing a single spatial frequency. The resulting high-resolution image is reconstructed from a series of nine sequentially recorded images corresponding to three linear translations of the pattern (phase stepping) and three rotation steps of 120° to obtain isotropic resolution enhancement. In such a scheme, spatial frequencies in the sample are downshifted by an amount equal to the spatial frequency of the striped illumination. As the excitation pattern is itself limited by the frequency passband of the imaging system, $2NA/\lambda$, spatial frequencies of twice this limit can be downshifted into the passband, leading to an approximately twofold increase in resolution.

A number of SIM variants have been developed which differ in the way patterns are generated and projected onto the sample. In the earliest implementations of SIM, illumination patterns were created with fixed diffraction gratings.[2] The excitation was diffracted into multiple beams which were recombined in the sample plane to form striped interference patterns. However, the fixed periodicity of the grating corresponds to a fixed spatial frequency. In practice, this means that the method could only be optimized for a single wavelength. For multi-color imaging this requires use of multiple grating, which is practically complex, or it must be accepted that the method will perform sub-optimally for other wavelengths. Pattern changing was achieved by mechanical rotation and translation of the diffraction grating, limiting operation at high speed. Later implementations make use of Spatial Light Modulators (SLMs) based on either liquid crystal technologies[3–5] or Digital Micro-mirror Devices (DMDs)[6–8] for pattern generation, which increases acquisition speed by an order of magnitude. However, these methods have their own limitations. For example, as is the case for diffraction gratings, new patterns have to be generated for every color to obtain the best resolution, and again channels must be recorded sequentially. Additionally, the discrete number and finite size of pixels in such devices restrict the periodicity of the patterns to integer numbers of pixels and introduce unwanted diffraction orders, which reduce photon efficiency.[3] Finally, all of these methods are highly sensitive on the polarization state of the incoming light, and this must be controlled in synchronicity with the pattern orientation to maximize pattern contrast in the sample.[9]

Here, we present an interferometric method to generate illumination patterns for SIM, MAI-SIM. The method makes use of an interferometer to generate fringes which are projected into the sample. The instrument projects patterns of optimal spatial frequency into the sample, irrespective of excitation wavelength. The method is fast, low in cost, and straightforward to implement. We demonstrate simultaneous SIM imaging in three colors over large fields of view (44 x 44 µm$^2$) at several frames per second recording speed. We maximize the potential of MAI-SIM with a suite of freely available software tools for image reconstruction. We call the method Machine learning Assisted Interferometric SIM, MAI-SIM, which is constructed using open design principles. All hardware and software modules to implement MAI-SIM are fully described, for anyone to implement on standard fluorescence microscopy frames. Finally, we demonstrate the power of MAI-SIM through application in live cell samples, where we image dynamic organelle-organelle interactions in three colors. MAI-SIM records, renders, and visualizes super-resolution output in real time.

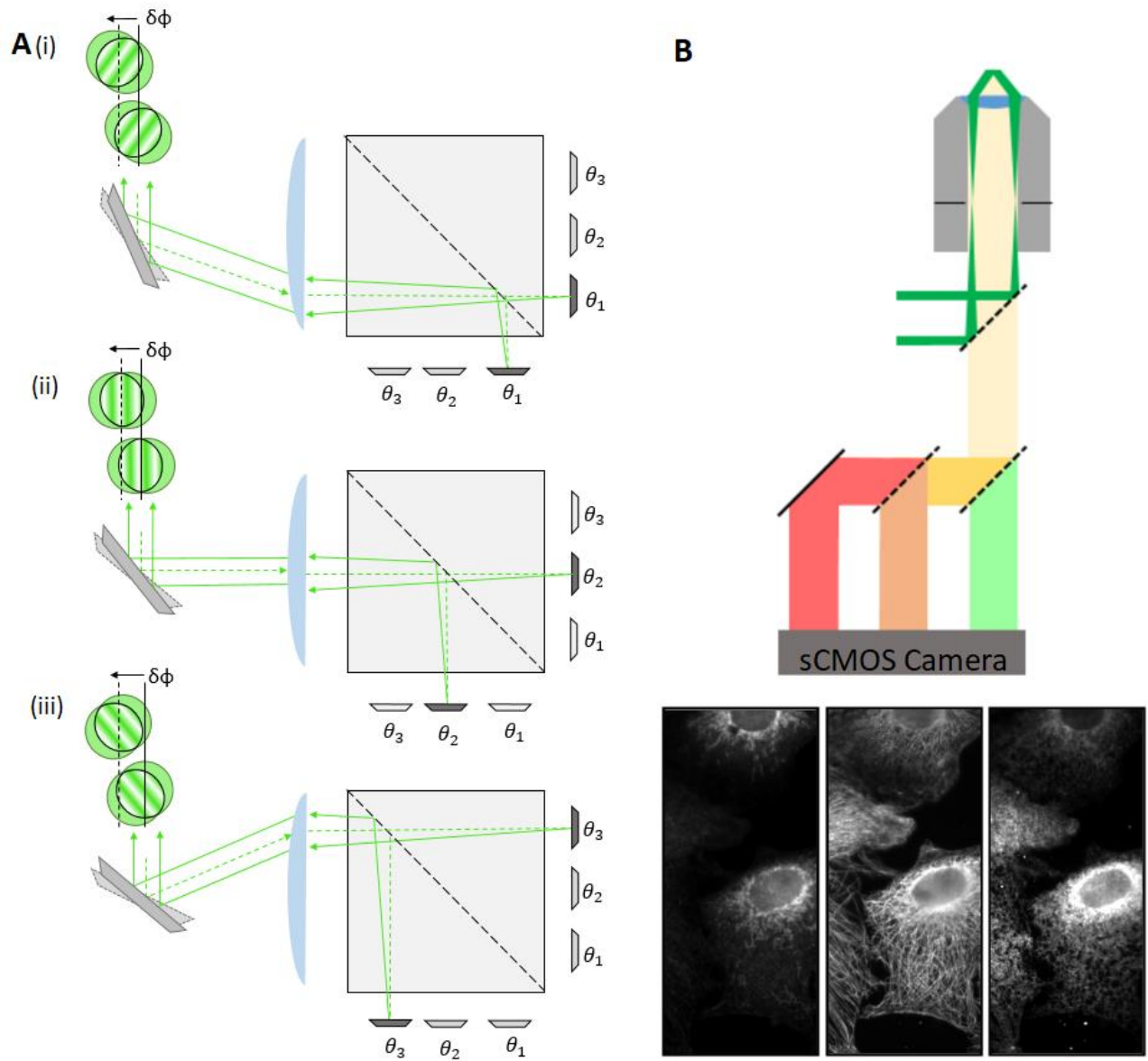

*Figure 1. Working principle of MAI-SIM. A(i-iii): Schematic of the interferometric design. Three small mirror pairs, each forming a Michelson interferometer, are tilted to form interference patterns at angles, $θ_{1-3}$, separated by $2π/3$. Large steps of the scanning mirror change the mirror pair addressed by the incoming beam (dashed line) and hence the rotation of the fringe pattern. Small steps of the scanning mirror laterally displace the position of the interference fringes and cause a phase shift, $δφ$, in the excitation pattern. B: Using several dichroic mirrors and emission filters, the three-way image splitter allows for the simultaneous display of up to three color channels on the camera chip. For full details on the optical setup and a rendering of the layout see SI Figure S1.*

## Results

### Hardware implementation

The core component of the presented microscope is a Michelson interferometer, where wavelength-dependent wedge fringes are projected onto the sample. A single high-speed mirror galvanometer is used to address one of three pairs of complementary 0.5" mirrors (Figure 1A). Each pair of mirrors forms fringes of constant thickness whose period is governed by the wavelength of the excitation light and the inclination of the mirrors with respect to one another. An image of these fringes, corresponding to an image of the surface of these mirrors, is subsequently relayed onto the sample (SI Figure 1&2). Furthermore, by adjusting the tilt of each mirror individually, fringes are produced such that moving from one mirror pair to the next rotates the fringe pattern around the optical axis. Thus, stepping from one mirror pair to another (Figure 1) rotates the pattern by 120°. Finally, phase shifting of the pattern is achieved using small galvanometer steps to translate the beam across each mirror pair, which yields a fixed lateral shift in the projected image plane. This implementation greatly reduces the complexity associated with interferometric system which typically require multiple moving comments which need to be precisely synchronised.[10,11] Because the fringes are coherently formed, it is critical that the coherence lengths of lasers used in the setup exceeds the maximum path difference between the interfering beams. Good temporal coherence is thus important, in contrast to SLM based SIM systems. In our setup, mirror pairs are on individual translation stages so that the path length difference can be precisely controlled to obtain optimal fringe contrast.

A key issue with SIM in high Numerical Aperture (NA) imaging systems is that the modulation depth of the pattern, which governs the strength of the super-resolution information collected, depends on the polarization orientation of the incident light. For high pattern contrast, s-polarization must be maintained for all pattern rotations. In many SIM systems, this is achieved through use of electro-optic liquid crystal devices which must be synchronized to the pattern generation process and the wavelength chosen. Here we chose to use a wedge polariser placed in the Fourier plane of the microscope as pioneered by Heintzmann et al.[12] To maintain the correct linear polarization at the back aperture of the objective, a matched pair of quad-band dichroic mirrors was used to compensate for the inherent ellipticity introduced by such mirrors. The detection path of the system was equipped with a commercial three-way image splitter, which included dichroic mirrors and emission filters to enable the simultaneous detection of three different fluorophores on a single high-speed camera (Figure 1B). The synchronization of the individual components is simpler than for many other SIM systems, as fewer active elements need to be considered.

**Instrument characterization**

To quantify the lateral resolution improvement, 100 nm fluorescent spheres were imaged at an excitation wavelength of 488 nm. We determined the full-width-half-maximum (FWHM) of the intensity profiles for individual point emitters by fitting the images to a 2D Gaussian profile (SI Figure S5). The mean FWHM of 70 emitters was measured as 200 nm for deconvolved widefield images and 120 nm for reconstructed SIM images, which is as expected for the spatial frequency of the fringe pattern. The performance of the system was tested using fluorescent microspheres with 200 nm diameter as well as mCherry-labeled tubulin, both of which have well-defined structures. The samples were imaged with an excitation wavelength of 561 nm, and reconstructed using an inverse matrix approach implemented in MATLAB.[13] Figure 2 compares the reconstructed SIM data with the corresponding Wiener filtered widefield images. While the latter provides a suppression of out-of-focus light and an enhanced visual appearance, SIM is capable of significantly enhancing the resolution, which is crucial to distinguish individual particles and filaments.

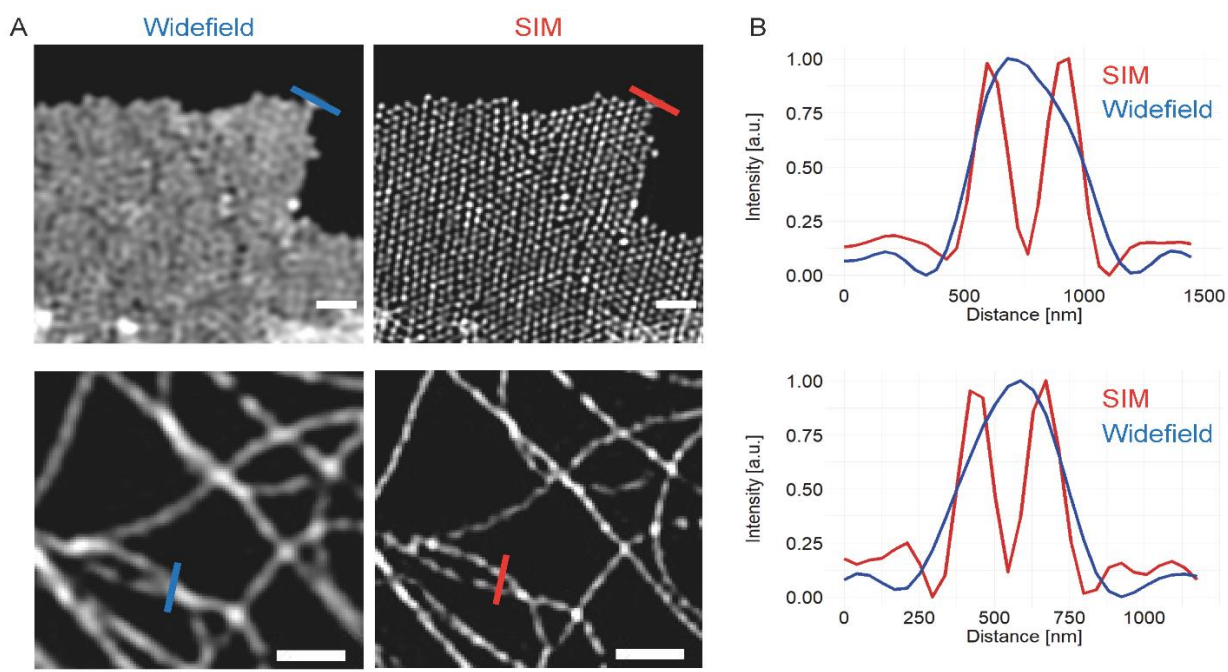

*Figure 2: Demonstration of the MAI-SIM on samples excited at a wavelength of 561 nm. In the deconvolved widefield images, adjacent 200 nm beads (top) and entangled microtubule filaments (bottom) remain unresolved, but are clearly distinguishable in the SIM image. Images were reconstructed using the inverse matrix method.[13] Scale bars are 1 μm.*

**The use of neural networks leads to enhanced image reconstruction in MAI-SIM**

In MAI-SIM, patterns of multiple colors can be projected into a sample plane simultaneously. Phase stepping is achieved through small lateral shifts across specific mirror pairs (see Figure 1A). As the fringe spacing is different for each wavelength, this means that not all colors will experience the same phase shift for the same lateral displacement. In practice, this means that the transverse pattern shifts required to reconstruct images may not correspond to the ideal 2π/3 phase steps. In sequential imaging this is not a problem, since the shifts can then be adjusted to be optimal for each individual wavelength. For the case of uneven phase steps, traditional Fourier based reconstruction techniques do not perform optimally.[10,13–16] We have compared the performance of several widely used reconstruction methods, specifically, the FairSIM-imageJ plugin[16], an inverse matrix method[13], an iterative blind SIM method[17], and finally our Machine Learning-SIM (ML-SIM) approach.[18] FairSIM is an easy-to-implement plugin with well-documented and versatile pre- / post-processing techniques[19] which are familiar to existing SIM users. In principle, FairSIM offers the capability of estimating phase step sizes from SIM patterns, however this capability only works for even phase steps. The inverse matrix method implements a more sophisticated method for phase estimation which is capable of performing reconstructions with uneven phase shifts. Its disadvantage compared to FairSIM is its limited versatility in pre- and post-processing modalities. Iterative blind SIM reconstruction is a fundamentally different approach and was developed for the reconstruction of SIM data acquired under random excitation patterns.[20–22] This makes the technique inherently agnostic to the specifics of phase steps used so long as a sufficiently good initial estimate for the sample image can be made.[17] For our ML-SIM, even steps are also not necessary and the method is specifically optimized for the reconstruction of interferometric SIM data, although it works equally well with all other implementations for SIM.

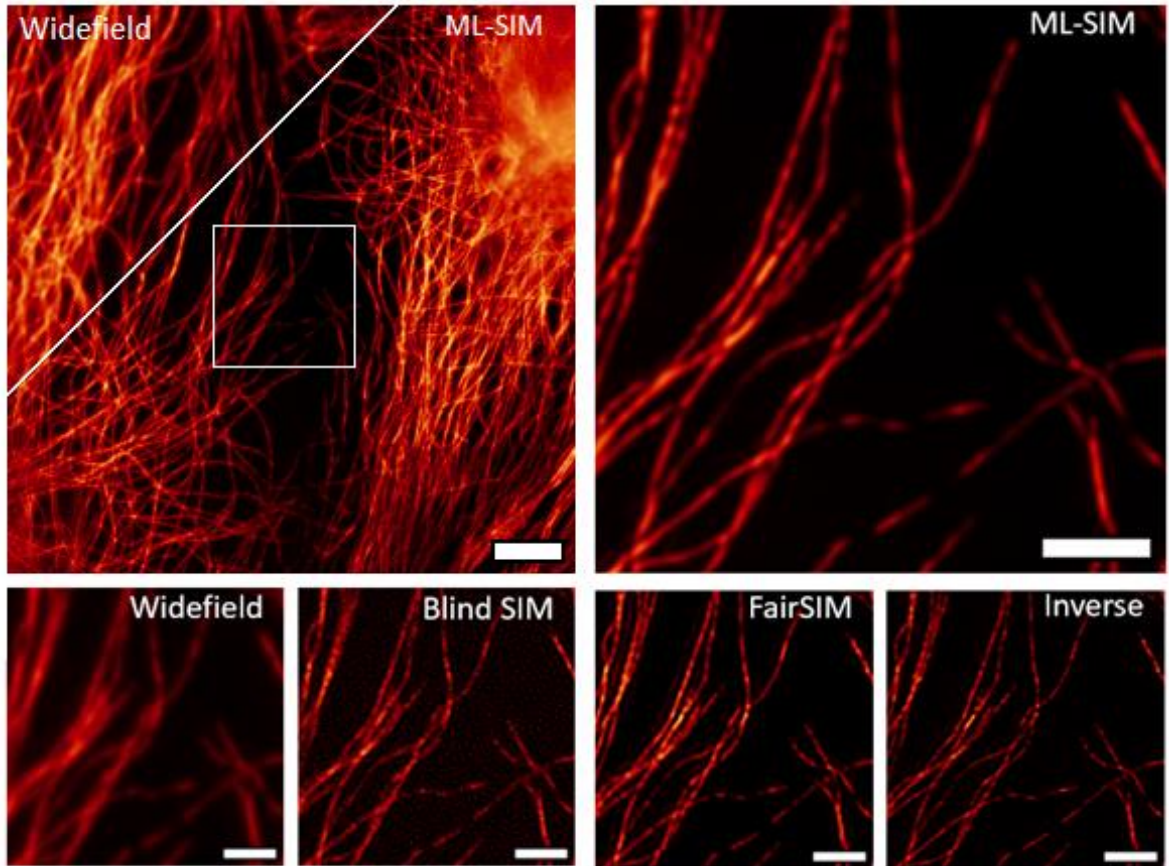

*Figure 3: Reconstructions of MAI-SIM data from fixed COS7 cells containing beta-tubulin labelled with secondary anti-bodies conjugated to ALexa568 dye. Top left: comparison of diffraction limited widefield images and ML-SIM reconstruction over the full FOV. Scale bar is 5 μm. Top right: ML-SIM reconstruction for magnified region indicated in the left-hand panel. Scale bar is 1 μm. Bottom, from left to right: wide field, blind SIM, FairSIM, and inverse matrix reconstructions over the same region. In all reconstruction methods, a resolution increase is apparent compared to the widefield image. In the case of ideal phase shifting, the FairSIM and inverse matrix reconstructions have performed comparably. The resolution increase achieved by the blind SIM method is less than that of the other techniques and there is significant amplification of background noise. Scale bars are 1 μm.*

ML-SIM is a deep neural network consisting of approximately 100 convolutional layers, similar in architecture to residual channel attention networks.[23] These are designed for single-image upsampling, where here the architecture is modified to take nine raw SIM frames as input from which the image is reconstructed. For this study, our previously published approach was developed to address the specific challenges of MAI-SIM. The model is trained on synthetically generated SIM data with a full physical model of the SIM process, which includes experimental factors such as spherical aberrations, Poisson noise, and finite pattern contrast. Data are simulated for various point spread functions, illumination angles, phase steps, and noise sources. The imaging parameters are randomized, such that the model learns to be robust irrespective of the specifics of a particular SIM imaging setup.

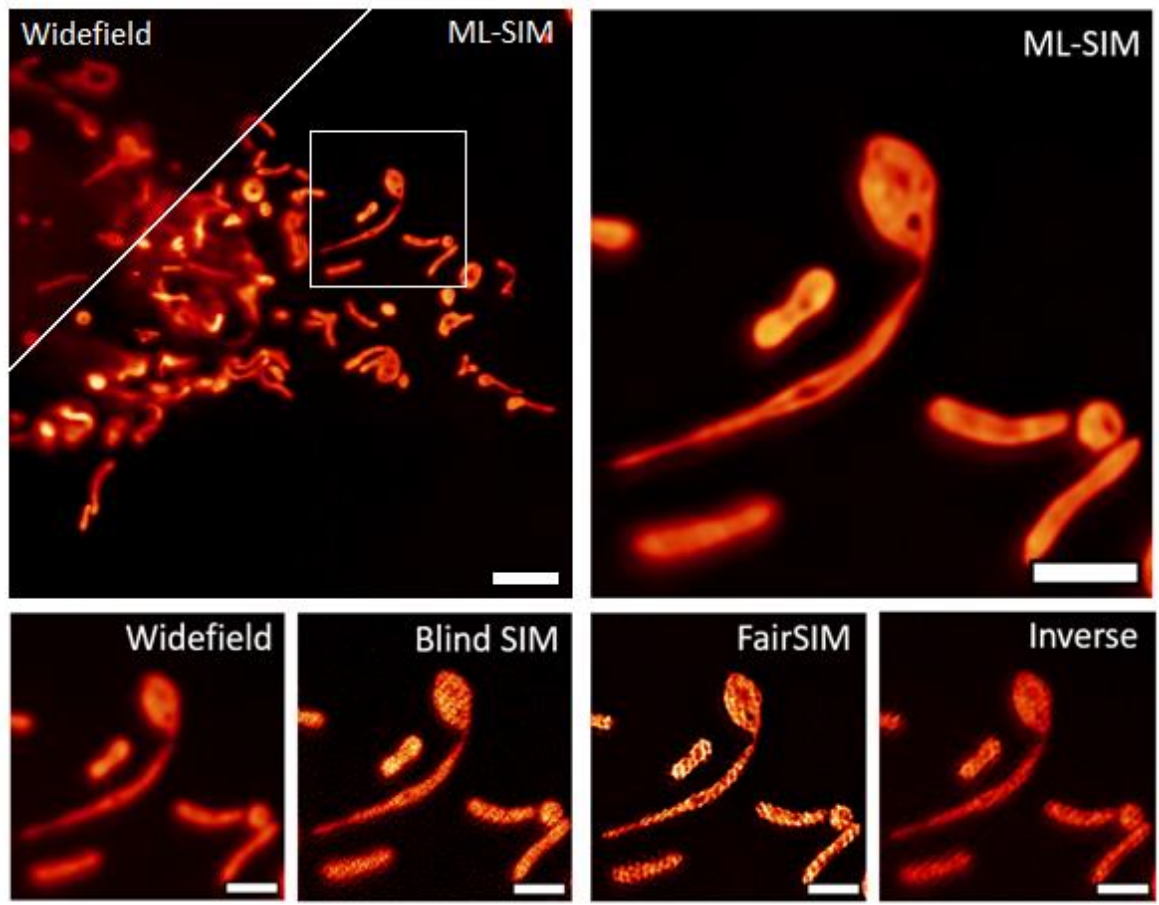

*Figure 4: ML-SIM outperforms classical SIM reconstruction algorithms for the analysis of data acquired with uneven phase steps. Reconstructions of SIM data from GFP-labelled mitochondria in live samples with data acquired during simultaneous three color imaging (only one color channel shown here). GFP was excited at 488 nm, but phase steps were optimized for the separate 561 nm color channel, leading to phase steps that differ from $2\pi/3$. Top left: comparison of diffraction limited imaging and ML-SIM over the full FOV. Both resolution and background rejection are enhanced in the ML-SIM reconstruction. Scale bar is 5 µm. Top right: ML-SIM reconstruction of magnified view indicated in left hand panel. The ML-SIM method is unaffected by the uneven phase stepping and resolution is improved without reconstruction artefacts. Scale bar is 1 µm. Bottom left to right: wide field, blind SIM, FairSIM and inverse matrix reconstructions for indicated region. The uneven phase stepping leads to significant reconstruction artefacts, including, striping artefacts and honeycombing. While the inverse matrix approach remains free from striping artefacts, additional honeycombing is apparent due to low signal to noise ratios. The blind SIM reconstruction shows a limited increase in resolution over the widefield image and amplification of background noise. Striping artefacts are also present as a result of a poor initial estimate of the sample. Scale bars are 1 µm.*

The quality of these reconstructions is compared in Figures 3 and 4. Figure 3 demonstrates the performance on interferometric SIM data from a sequential acquisition of multiple color channels. This permits the phase shifts to be optimized individually for all wavelengths. As expected under these conditions, the performance of the inverse matrix and FairSIM is comparable, because the assumption of equidistant phase steps of $2\pi/3$ is valid. In contrast, while a resolution improvement over the widefield image is still apparent, the blind SIM approach introduces noticeable reconstruction artifacts, associated with the amplification of noise. Figure 4, on the other hand, compares the performance of these reconstruction techniques on data acquired during simultaneous acquisition of all color channels. Here, the phases were optimized for 561 nm excitation, i.e. with equidistant phase steps separated at 2π/3, however the images shown are for the 488 nm excitation channel, where phase steps are no longer equidistant and differ significantly from 2π/3. As seen, ML-SIM outperforms the classical reconstruction methods by a considerable margin. The algorithm handles uneven phase steps and deals better with data acquired at low Signal-to-Noise Ratios (SNR). For the imaging of dynamically changing and light sensitive samples, where short camera exposure times and low excitation intensities have to be used, this presents very significant advantages. We quantified the relative performance of the different methods using NanoJ analysis software[24] (SI Figure S8, and SI table 1).

In addition to an increased robustness to noise and pattern phase errors, the use of machine learning for SIM enables image reconstruction that is considerably faster than Fourier or iterative methods. We leverage this advantage to demonstrate on-the-fly reconstructions of live generated SIM data. On an entry-level graphics card, image acquisition, reconstruction, and visualization of up to three color channels can be performed simultaneously at a frame rate of 1 Hz over a 44 x 44 μm field of view (SI Figure S8). Video 1 in the SI presents a demonstration of live MAI-SIM imaging, demonstrating the responsiveness and ease of use of the method, even for non-experts.

**MAI-SIM enables imaging in multiple color channels**

To assess the performance of MAI-SIM for imaging in multiple colors, we first imaged labeled fluorescent microspheres and intracellular organelles in fixed COS-7 cells. Figure 5 shows multicolor microspheres excited at 488 nm, 561 nm and 647 nm, respectively. Data in the three channels were acquired sequentially so that phase steps could be optimized for every color. In this case, classical reconstruction algorithms work well. Images shown were reconstructed using the inverse matrix method and are clearly better resolved than the deconvolved widefield images shown for comparison. Figure 5B shows fixed GFP expressing mitochondria (blue), fixed mApple stained endoplasmic reticulum (ER, yellow), and fixed microtubules labeled with Alexa647 immunostaining (red). In all three images, filament networks, bundles, and junctions can be resolved at sub-diffraction resolution. This demonstrates that artifact-free, high-resolution SIM can be obtained for a variety of samples and labeling methods with classical reconstruction methods.

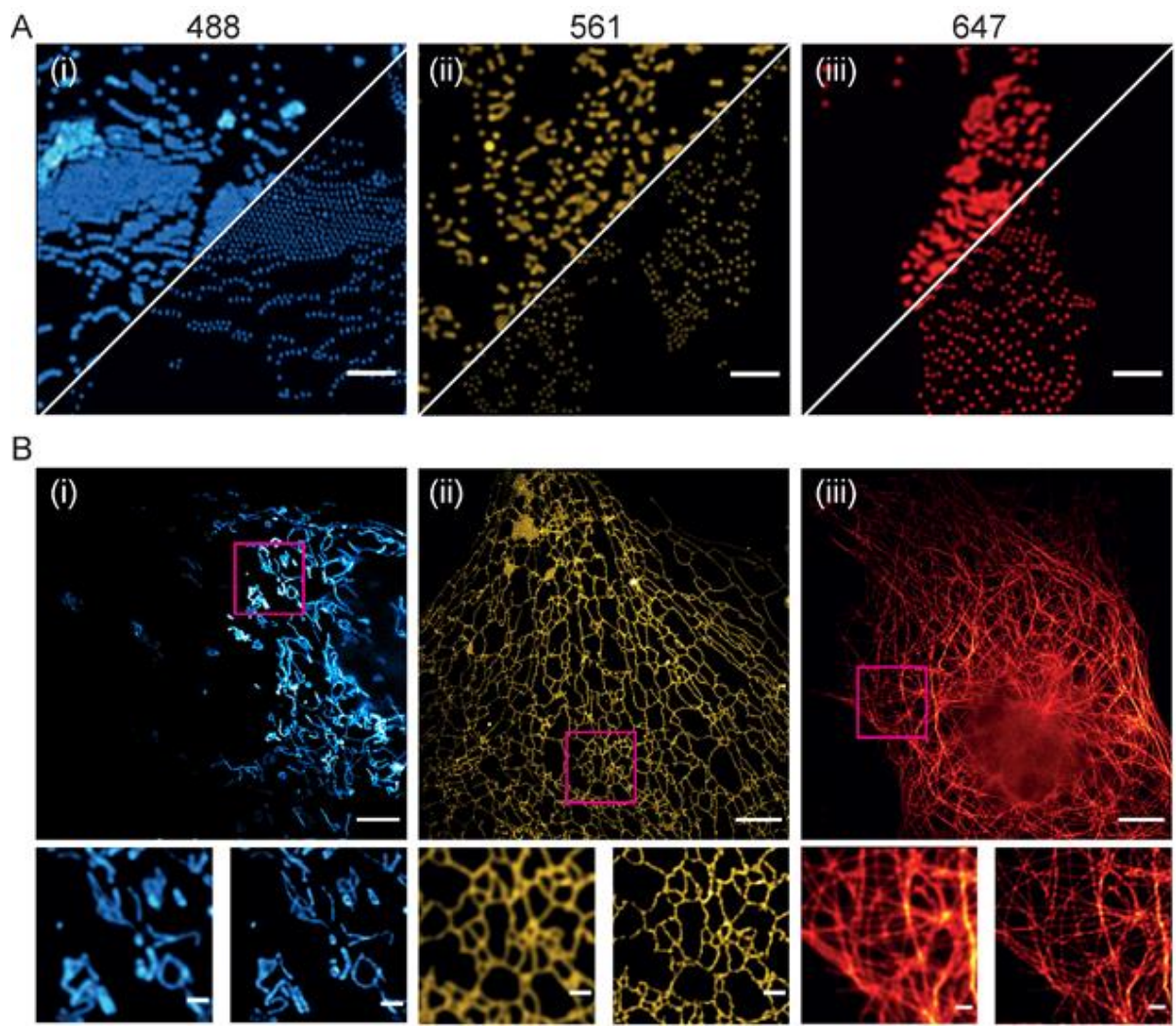

*Figure 5. Multi-color imaging capability of the developed instrument. A: 200 nm TetraSpeck microspheres imaged with excitation/emission wavelengths of 488/515 nm, 561/580 nm, and 647/680 nm, respectively. In each panel, the top left corner shows the deconvolved widefield image, and the lower right corner shows the inverse matrix SIM reconstructions. Scale bars are 2 μm. B: Multicolor imaging of cellular organelles with different staining strategies. (i) Fixed mitochondria (GFP), (ii) fixed ER (mApple), (iii) fixed and immunostained tubulin (Alexa647). Scale bars are 5 μm (top/middle) and 500 nm (bottom).*

Finally, we assess the use of MAI-SIM for the investigation of live biological samples. The ability to image multiple color channels simultaneously without temporal offset is crucial for the investigation of dynamic interactions between different biological structures, yet is not available with existing SIM modalities. To demonstrate this capability, live COS7 cells were labeled with mitotrackerGreen, mApple-Sec61b-C1 conjugate, and SiR-lysosome. Figure 6 shows a visualization of mitochondria-ER interactions during ER reorganization. The simultaneous acquisition of SIM frames in the different color channels permits the observation of ER structures wrapping dynamically around the mitochondrial periphery without any temporal offsets. The arrow indicates the location of a contact site between the ER (magenta) and a mitochondrion (cyan). As the mitochondrion moves, contact between the two structures is seen to be maintained, an observation that could not have been made with sequentially performed SIM imaging (see video 2 in SI).

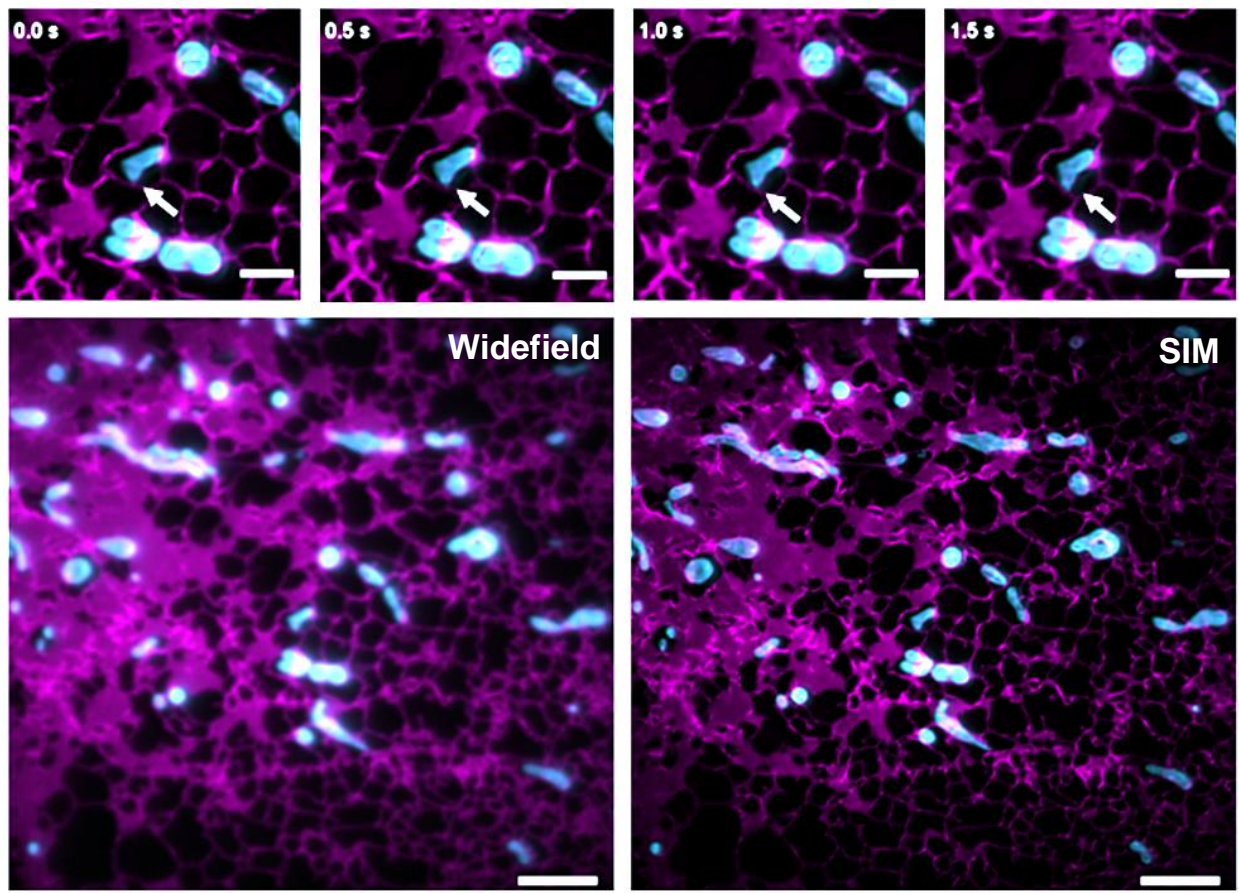

*Figure 6. Simultaneous imaging of live cells in multiple color channels. COS7 cells were stained with mitotrackerGreen and transfected with mApple-Sec61b-C1 to label ER. Top panel: Timelapse images of mitochondria embedded in rapidly reorganizing ER networks. The arrow indicates a contact site between the mitochondria and the ER. Contact is seen to be maintained as the structures reorganize. Images were taken at 0.5 s intervals. Scale bar is 2 $\mu$m. Bottom: Comparison of diffraction-limited (left) and SIM (right) images. Scale bar is 5 $\mu$m. Simultaneous imaging of multiple channels permits interactions between rapidly moving small structures to be investigated without temporal (and thus spatial) offsets.*

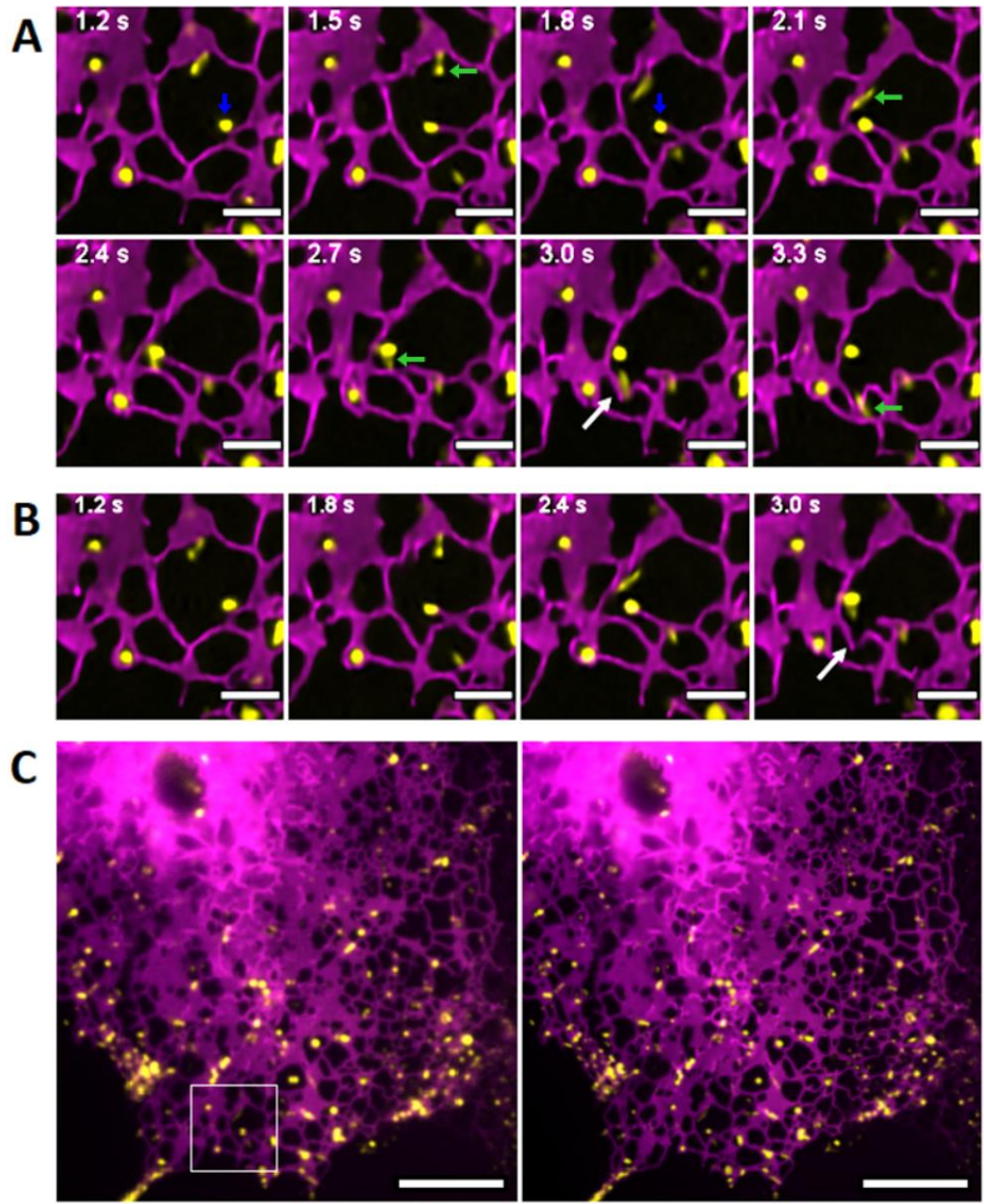

*Figure 7. Simultaneous imaging with MAI-SIM improves temporal resolution and image reliability. COS7 cells stained with SiR-lysosome (yellow) and transfected with mApple-Sec61b-C1 to label ER (magenta). A: Time series of two ER network reorganization events visualized with MAI-SIM. One lysosome (blue marker) can be seen to pull a new filament from the network which fails to make a connection and is subsequently released. A second lysosome (green marker) moves rapidly along the ER network and cleaves an ER filament (white arrow) to form two new filaments. Scale bar is 2 µm. B: The same ER network events visualized in a sequential SIM acquisition. In this case the temporal offset between the color channels means that the fast cleaving event (white arrow) appears independent of lysosome transport. Scale bar is 2 µm. C: Full field of view of ER network in widefield (left) and MAI-SIM (right) imaging. The region showing the time series data is indicated by the white square. Scale bar is 10 µm.*

It has recently been suggested that lysosomes play an active role in the rapid reorganization of ER network topology, providing the pulling force for rapid tubule formation and the forming of ER network connections.[25] Figure 7 shows the unique capability of MAI-SIM to track such dynamic cellular processes. In the images shown, lysosomes appear in yellow, and the ER network in magenta. Using MAI-SIM with the detection splitting optics (Figure 7A) leads to perfect co-registration of the two color channels in time, and an effective doubling of the achievable frame rate with respect to sequential SIM imaging. The bottom row shows the same event, but this time using the camera frames in the data that mimic a sequentially recorded SIM image (Figure 7B). The camera frame rate was 20 frames per second over a 44 x 44 µm$^2$ field of view; the panels shown in the sequences are cropped regions from the full frame view (Figure 7C). Two events are seen taking place. As demonstrated in Figure 7, a lysosome (indicated with the blue marker) is seen to pull an ER tubule towards another region of the ER but ultimately fails to make a connection, the lysosome detaches, and the tubule subsequently retracts. A fast moving lysosome (green marker) is seen to travel downwards, cleaving an existing tubule (white arrow) to form two new tubules. Both of these events would have been hard to visualize or interpret on sequentially recorded SIM data. We simulated sequentially recorded SIM by analyzing the nine frames required to generate a SIM image for the lysosome channel and the following nine frames for the ER channel. The sequencing of the frames thus simulates a traditional SIM experiment and leads to a halving of the effective frame rate. The corresponding data are shown in Figure 7B. Neither of the two phenomena described above could have been correctly interpreted in the sequential data, because the correlation between the organelle locations is lost. MAI-SIM is thus an ideal technology to study organelle biology such as these transient interactions between lysosomes and the ER.[25]

**Discussion**

We have developed a fast, structured illumination microscope which uses 2 beam interferometry to generate illumination patterns. The method is ideally suited for the investigation of fast biological processes in multiple color channels. The pattern generation process is optimal for any color, and patterns can be projected into the sample simultaneously. Combined with a multichannel image detection system, this allows for sub-diffraction resolution imaging in multiple colors simultaneously while maintaining optimal resolution for all excitation wavelengths.

The process is straightforward and cost-effective to implement on existing widefield microscope frames. There are unique aspects to interferometric SIM which make the use of traditional reconstruction algorithms challenging. We address this problem by modeling the physical process of interferometric SIM to generate *in silico* data that can be used to train a neural network. The hardware and software together provide a powerful and easy-to-use high speed SIM platform.

Our approach allows for the reliable reconstruction of low SNR SIM images in which pattern phase shifts cannot be precisely controlled, for example during simultaneous data acquisition in multiple color channels where the system has been optimized for the central excitation wavelength. The analysis package is also applicable for use in traditional

SIM modalities but has the added advantage of speed. Once the network is trained, reconstruction and visualization can be achieved in multiple colors frame rates exceeding 1Hz (higher for smaller fields of view). MAI-SIM data is acquired with no temporal offsets between the channels, enabling the observation of organelle-organelle interactions in real time. Live analysis and visualization of MAI-SIM data is possible. We demonstrate the method in the study of interactions between mitochondria, ER and lysosomes, and highlight how their interactions control and affect ER remodeling processes. The presented system is compact and, with only one actively moving component, does not require complex synchronization procedures, electronic control of the polarization state, or use of expensive and photon-inefficient spatial light modulators. MAI-SIM could easily be extended to permit TIRF-SIM imaging[26] and an extension of the method to enable 3D SIM would be possible through the addition of a third arm to the interferometer (see SI Figure S8 for a proposal on how this might be achieved).[27,28] The microscope is suitable for use in a variety of biological applications and is a powerful tool for multicolor imaging of living cells and intracellular organelle interactions in real time. We therefore envisage MAI-SIM to be an attractive alternative to commercial and SLM-based SIM instruments, making high-resolution imaging available for a wide scientific community.

## Methods

Methods, including statements of data availability and any associated accession codes and references are available in the online version of the paper. The code used for hardware control and data processing and live ML-SIM can be found at the GitHub repository: https://github.com/edward-n-ward/MAI-SIM

## Acknowledgements

The authors would like to acknowledge James Manton and Florian Ströhl for their helpful insights and contributions to the design.

## Author contributions

ENW and LH designed and constructed the system, developed the image processing software and collected data. CNC, ENW and JRL developed the machine learning reconstruction. AW and JRL contributed to the design of the setup. AW and ENW developed the iterative reconstruction method. CWC, LM, and ML prepared fixed and live biological samples. CFK and CJR conceived the interferometry method. CFK and GSK supervised the project. ENW, LH, and CFK wrote the manuscript.

[This page intentionally left blank]

# MAI-SIM: interferometric multicolor structured illumination microscopy for everybody – Supplementary information

Edward N. Ward, Lisa Hecker, Charles N. Christensen, Jacob R. Lamb, Meng Lu, Luca Mascheroni, Chyi Wei Chung, Anna Wang, Christopher J. Rowlands, Gabriele S. Kaminski Schierle, Clemens F. Kaminski

| | |
|---|---|
| **Supplementary Figure 1** | **Optical path for fringe generation** |
| **Supplementary Note 1** | **MAI-SIM Hardware** |
| **Supplementary Figure 2** | **Influence of path length on modulation depth** |
| **Supplementary Figure 3** | **Optimization of pattern orientation** |
| **Supplementary Figure 4** | **Pattern phase optimization** |
| **Supplementary Figure 5** | **Effects of phase shifting on reconstruction quality** |
| **Supplementary Note 2** | **Reconstruction methods** |
| **Supplementary Figure 6** | **Resolution estimation on single beads** |
| **Supplementary Figure 7** | **Error map comparison of reconstruction quality** |
| **Supplementary Table 1** | **Reconstruction quality measured by RSP and RSE** |
| **Supplementary Figure 8** | **Live image reconstruction with machine learning** |
| **Supplementary Note 3** | **Sample preparation** |
| **Supplementary Note 4** | **Extension to 3D-SIM** |
| **Supplementary Figure 9** | **Proposed optical setup for 3D MAI-SIM** |

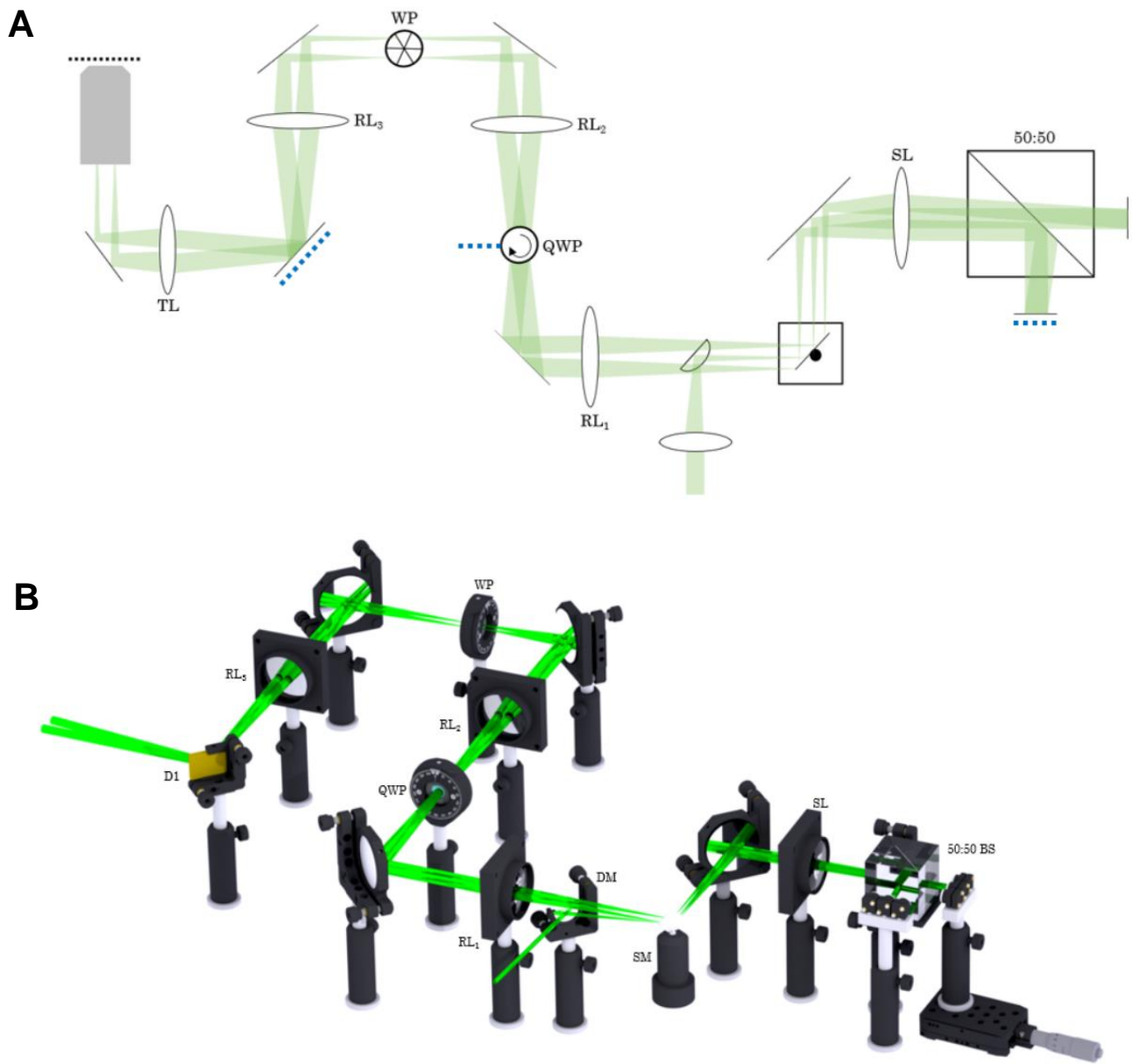

**Figure S1: Optical path for fringe generation**

Simplified optical schematic (A) and rendering of typical setup (B) for MAI-SIM pattern generation. Three laser lines are used in the MAI-SIM system with wavelengths of 491 nm (Cobolt Calypso, 200 mW), 561 nm (Oxxius SLIM-561-150, 500 mW) and 647 nm (Toptica iBeam SMART, 100 mW). These are combined coaxially using dichroic mirrors. The beam profiles are cleaned using a pinhole at the focus of a Keplerian beam expander to provide uniform illumination across the whole FOV. The expanded beam is focused onto the scan mirror (**SM**) (Scanlab, dynAXIS-M) using a 150 mm focal length lens (Thorlabs, AC254-150-A) and a D-mirror, **DM**, (Thorlabs, PFD10-03-F01) to separate the incoming and outgoing beams of the interferometer. After reflection by the scan mirror, the beam is re-collimated by a scan lens, **SL**, (Thorlabs, AC508-150-A) and subsequently enters the interferometer. A 2-inch 50:50 beam splitter cube (Thorlabs, BS031) is used for amplitude splitting and the two resulting beams each hit one of two complementary mirrors in an arrangement of 3 pairs of individually adjustable ½-inch mirrors (Thorlabs, BB05-E01). Complementary mirrors are tilted with respect to another to generate wedge fringes of the correct orientation. The small mirrors are additionally placed individually on micrometer translation stages (Thorlabs, XRN25P / XRN25C) and This enables the adjustment of the path length difference for each interferometer configuration individually to achieve optimal contrast in all orientations (SI figure 3).[1] The two beams reflected from one of the pairs pass back through the beam splitter and the scan mirror. Both beams then pass over the top of the D-mirror and are combined by a series of 150 mm focal length lenses, **RL$_{1-3}$**, (Thorlabs, AC508-150-A) to form an interference pattern at an angle determined by the scan mirror position and corresponding angle of small mirror Main text (figure 1A). The beams are then relayed through a wedged polarizer, **WP,** (CODIXX, custom part). This static element ensures the polarization is correct for all orientations, while maintaining the multicolor capabilities of the system.[2] An achromatic quarter waveplate, **QWP**, placed in the image plane of the relay is used to generate circularly polarized light before the beam hits the wedged polarizer. This ensures equal intensities for all fringe orientations. The beams enter the inverted microscope frame (Olympus, IX73) via a quad-band dichroic mirror, **D$_1$**, (Chroma, ZT405/488/561/640rpcv), which is from the same production batch as the dichroic mirror mounted in the microscope frame. The arrangement separates excitation and emission light paths. Pairing these mirrors from the same batch ensures that any ellipticity introduced by birefringence of one mirror is cancelled out by the other. Thus s-polarization can be maintained for all orientations. Similar consideration apply to other conjugate mirror pairs in the optical setup.[3] The beams are focused onto the back focal plane of the objective lens (Olympus, UPLSAPO60XW) by 300 mm focal length tube lens, **TL**, (Thorlabs, AC508-300-A) and interfere in the sample plane to form a sinusoidal illumination pattern. Intermediate image planes (i.e. the location of fringe formation) are shown with blue dashed lines. The resulting fluorescent signal is collected by the same objective and passes through the dichroic mirror, onto an image splitting device (Cairn, Optosplit III). The image splitter consists of 2 dichroic mirrors and corresponding passband emission filters to separate individual color channels according to the chosen fluorophores. Up to three channels are then displayed side by side on a sCMOS camera with 2048 x 2048 pixel chip and 6.5 x 6.5 µm² pixel size (PCO edge 4.2 bi).

Synchronization in the final system is achieved by controlling the camera exposure, stage control, and mirror movement through a LabVIEW interface, while the image data was read in a separate application (Micromanager).[4] Digital I/O lines from a DAQ card (National Instruments, BNC-2110) were used to control the exposure time and monitor the data readout from the camera using the I/O lines from the camera. In parallel, the mirror position was controlled using the analogue I/O lines from the same DAQ card. We found that the time taken to move and settle the mirror at different positions is comparable to the data readout time for a 512x512 pixel image. Frame rate was therefore maximized by moving the mirror after the exposure had finished while the data were read from the camera. Small digital delays were added before and after mirror movement ensure that the mirror was stationary during camera exposure. The stage was controlled using serial commands in the same LabVIEW interface. Stage movement was similarly synchronized to camera readout to maximize framerates although extra digital delays were needed before the following acquisition as z-movements were significantly slower than camera readout. All digital delays (see timing diagram on project [page](page)) were determined empirically by imaging a thin fluorescent layer and verifying that stripe patterns were steady during exposures. It was found that settle times depended heavily on the mirror step distance and longer settle times were after changing fringe orientation than after phase stepping.

# Supplementary Note 1: MAI-SIM Hardware

**The Michelson interferometer:**

The Michelson interferometer is a commonly used amplitude splitting interferometer which, due to its simplicity and high accuracy, is popular in a wide range of applications.[1] In such a system, a beam splitter divides a primary wave into two beams which are subsequently reflected by two mirrors in the corresponding optical path. The two returning beams interfere after being recombined by a condensing lens and create periodic intensity fringes, when incident on a screen or detector. A tilted mirror pair leads to the formation so called fringes of constant thickness. The angle of mirror inclination determines the frequency of the pattern, while the path length difference (among other factors) defines the pattern contrast for non-monochromatic sources. For a clean sinusoidal fringe pattern with a high contrast, good quality light sources with high thermal stability and a narrow bandwidth are essential. Temporal coherence of the two interfering beams is important for the pattern quality as high contrast interference patterns will only be visible if the optical path length difference (i.e. twice the wedge thickness) between the interfering beams is smaller than the coherence length of the light source. The alignment of the interferometer mirrors is greatly facilitated when lasers coherence lengths of several mm or more are used.

To adjust the path length differences and optimize the fringe contrast in the MAI-SIM system, a camera was mounted in an intermediate image plane to directly observe changes to the illumination pattern. The optical path length difference was then adjusted for each chosen set of mirrors, using the micrometer translation stages on which one of the mirrors was mounted.

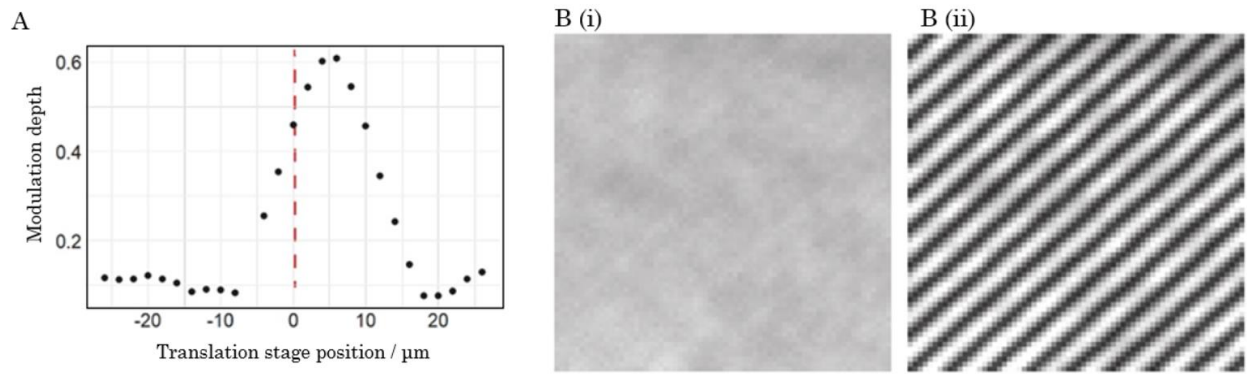

*Figure S2: Dependency of pattern frequency for low coherence length 647nm diode laser. A: Plot of fringe contrast measured on a camera in an intermediate image plane as a function of translation stage position. C: Images acquired on the intermediate camera with un-optimized (i) and optimized (ii) stage positions.*

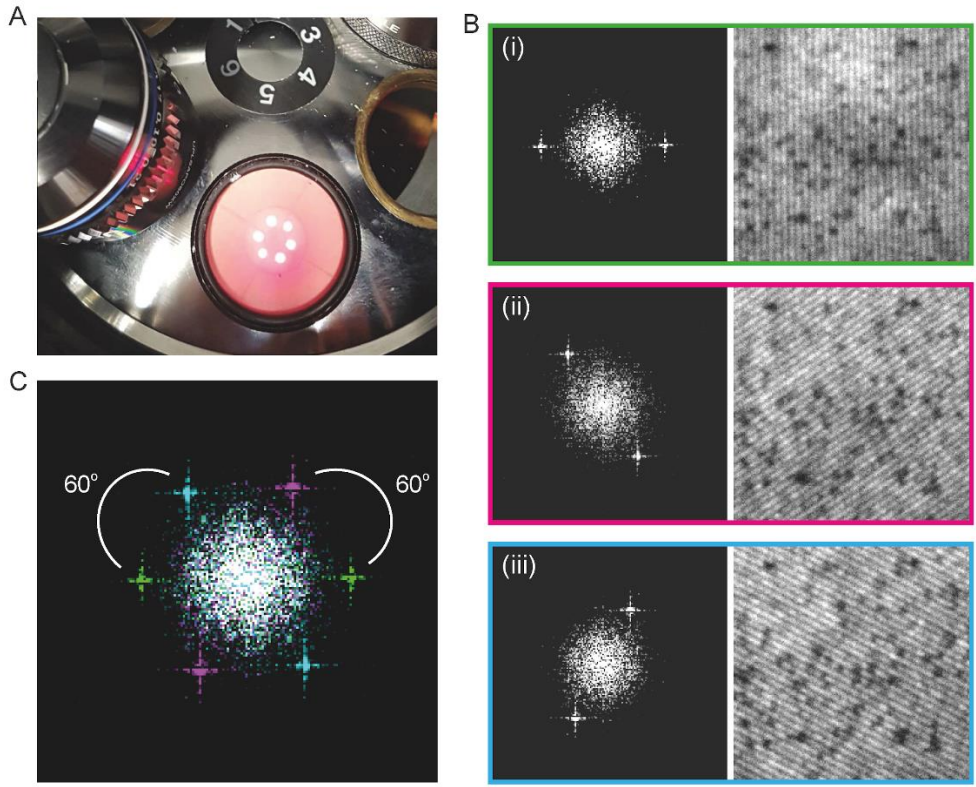

*Figure S3: Positioning of the excitation spots at the back of the objective. A: An alignment target is used to provide an approximate estimate of mirror positions. B i-iii: The fringes are then viewed on a spincoated fluorescent layer of 100 nm beads and the Fourier transform of the acquired image is then used to determine an appropriate fringe spacing. C: The pattern rotation is optimized using the average Fourier transform for all pattern orientations.*

**Pattern alignment:** Pattern orientation and period are controlled by directing the beams into the appropriate positions at the back of the objective by adjusting the three pairs of small mirrors. For precise alignment, a fluorescent bead monolayer was viewed and the stripes brought into focus (Figure S3). A live FFT implemented through MATLAB was then used to view the frequency components of the stripes in real-time, permitting precise alignment of the spots in the back focal plane of the objective.

**Pattern orientation change and phase stepping:** The system utilizes one scan mirror to achieve both rotation of the the fringe pattern in the sample (by stepping from one mirror pair to the next) and phase stepping (by small angular displacements). To determine the step size required for adequate phase shifts, a camera is placed in the intermediate image plane, once the spot position and pattern frequency have been optimized. This enables a direct observation of the pattern and voltages supplied to the scan mirror are adjusted until a phase shift of approximately 2pi/3 is observed on the camera. After thus coarse visual adjustment, a finer adjustment is achieved by summing the images from the three phase steps and confirming a cancellation of the stripes in the resulting image, which should be uniform in intensity (SI figure 5). If sequential imaging of different color channels is to be performed, this process can be optimized for each color separately. For simultaneous multi-color in conjunction with the image splitter, the central wavelength (in our case 561 nm) is chosen to optimize the phase steps. While this enables fast imaging, the phase steps will no longer be 2pi/3 for the other colors, and inverse or Fourier based reconstruction methods may no longer work leading to reconstruction artefacts (figure S6). As an additional control correct pattern shifting, the process described above is repeated by imaging the pattern shifts in the actual sample plane.

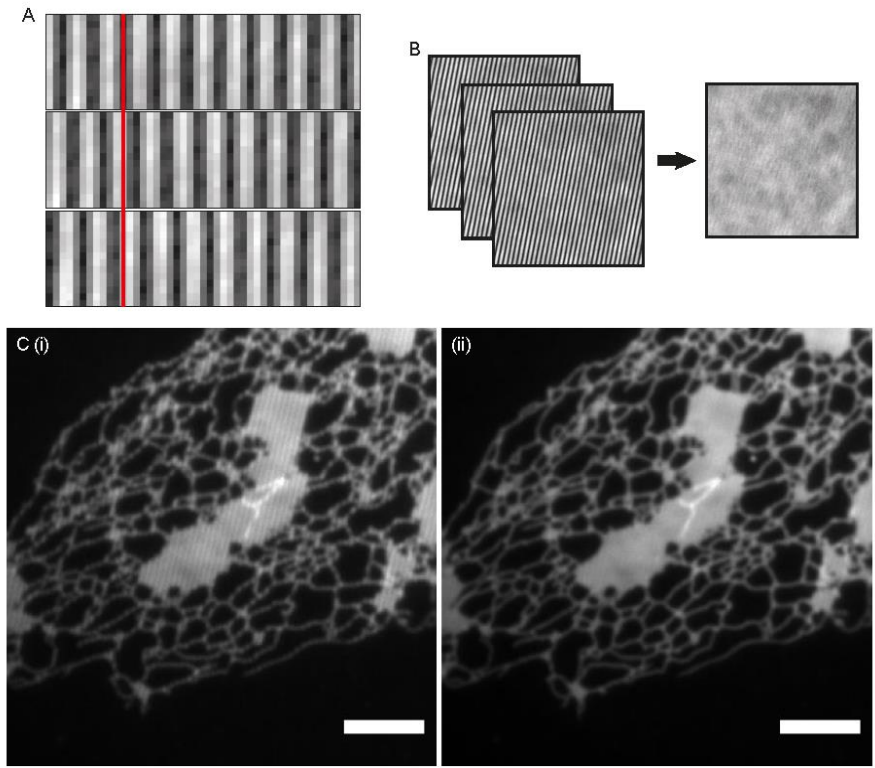

*Figure: S4 Pattern phase optimization. A: fringes formed by the 561nm laser viewed on the intermediate camera. B: After correct selection of scan mirror voltages, the phases were checked by summing images acquired for all three pattern phase shifts to produce even and flat illumination. C: Widefield images of mApple labelled ER in COS7 cells produced by summing images acquired at three pattern phases. Poorly (i) and well (ii) adjusted phase voltages can be confirmed by checking for residual pattern modulation in the widefield image. Scale bar is 5μm*

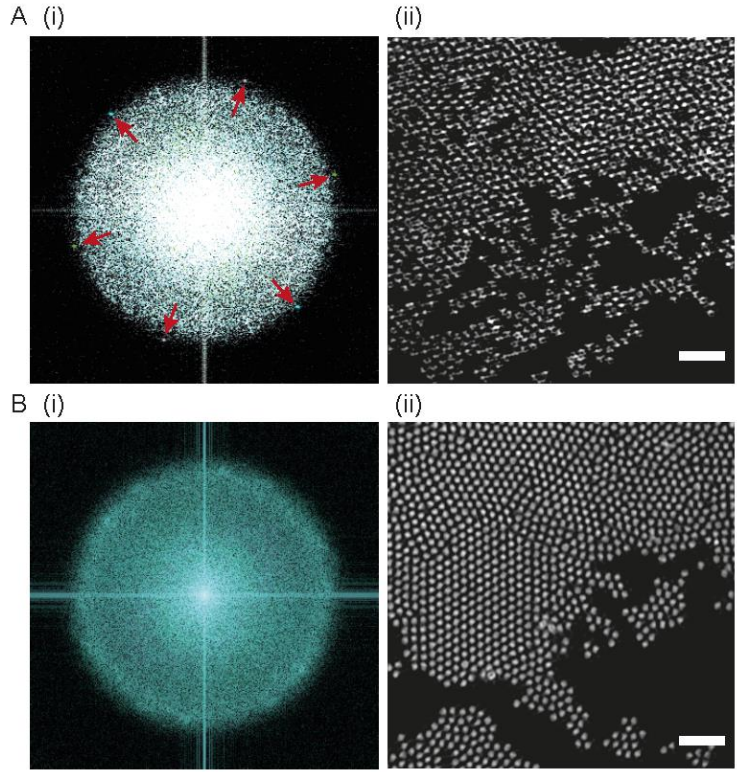

*Figure: S5 Results of pattern phase optimization on image reconstruction*. A i: Fourier spectra of the widefield image taken as the average of the 9 raw SIM frames. Peaks can be seen corresponding to stripes on the image. A ii: SIM reconstruction performed with FairSIM on SIM images with un-optimized phase steps. Uneven phase stepping results in significant reconstruction artefacts. B i: Fourier spectra of the widefield image taken as the average of the 9 raw SIM frames. No residual peaks can be seen meaning a correct 2pi/3 phase step has been achieved for all pattern orientations. B ii: SIM reconstruction performed with FairSIM on SIM images with optimized phase steps. Images of 200nm beads were acquired upon excitation with the 561nm laser line. Scale bar is 1µm

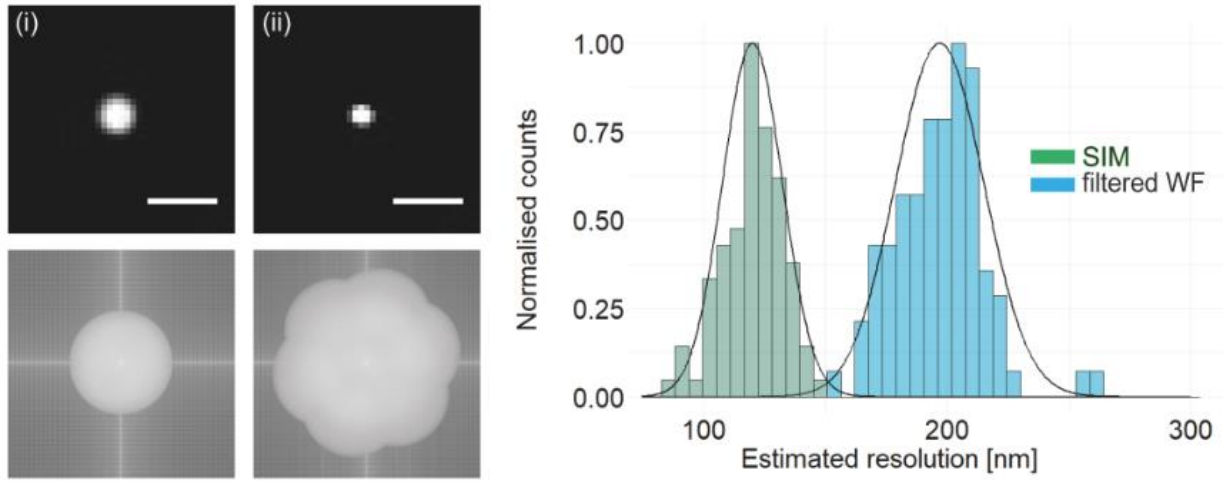

*Figure S6: Resolution enhancement on single beads.* The resolution improvement provided by the instrument is quantified using 100 nm fluorescent microspheres excited by the 488 nm laser line. Top: the particle size in SIM (right) is significantly reduced compared to the widefield image (left). Scale bars 500 nm. Bottom: the measurement of 70 particles suggested a mean diameter of 120 nm for particles imaged with SIM, while spheres in widefield mode have a diameter of 200 nm.

**Detection optics and channel registration**

For a coarse alignment of the detection system and channel registration, a ruled grid slide (Thorlabs, R1L3S2P) was illuminated with the brightfield lamp of the system, brought into focus, and centered on the camera. The filter cubes of the image splitter, containing 2 dichroic mirrors and 3 emission filters, were inserted and the adjustable mirrors in the device were aligned such that the same region of the slide was visible in all three color channels. Fine adjustment of the channels was performed before each measurement. The required dichroic mirrors and emission filters were inserted into the filter cubes according to the fluorophores in the sample. Fluorescent 100 nm diameter which emit into all three color channels were imaged onto the camera and the mirrors in the image splitting devices were adjusted for a chosen cluster of beads to line up in the same position in all channels. Images were then recorded in two or three colors simultaneously and fine registration was performed after acquisition using a calibration image of a sparse 100 nm bead sample.

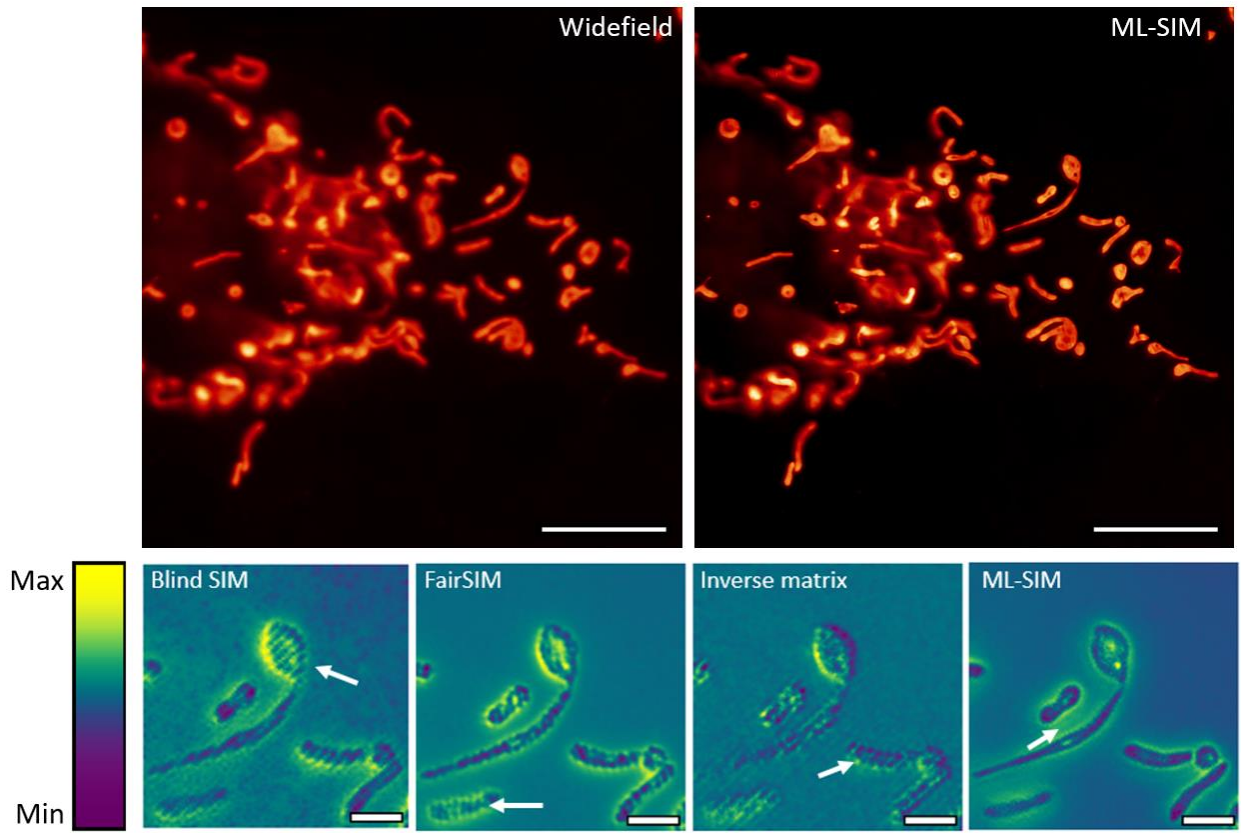

*Figure S7: Error map comparison of MAI-SIM reconstruction techniques.* Top: Comparison of widefield and ML-SIM reconstructions of MAI-SIM data acquired of GFP labelled mitochondria in live COS7 cells. Pattern phase shifts were optimized for 561 nm excitation while imaging was performed with 488 nm excitation, leading to uneven phase stepping. Scale bar is 10 µm Bottom: Normalized reconstruction error maps for figure 4 (main text) calculated using NanoJ SQUIRREL. From left to right; Blind SIM: diagonal striping artefacts (white arrow) are present in the image. As blind SIM does not require specific illumination patterns, these artefacts can be attributed to an improper initial estimate of the sample, indirectly caused by irregular phase shifts. FairSIM: honeycombing and vertical stripe artefacts are present (white arrow) which result from irregular phase shifts, poor estimates of pattern frequency and reduced modulation depth. Inverse matrix: As with Blind SIM, the principal artefact is diagonal striping (white arrow). In the case of the inverse matrix reconstruction, this is a result of pattern phase shifting being irregular and beyond what the algorithm is capable of handling. ML-SIM: the error in ML-SIM reconstructions is primarily limited to the background (white arrow) where out of focus signal dominates and not the key sample structures. Scale bar is 2 µm

# Supplementary Note 2: Reconstruction methods

During high-speed multi-color imaging, low signal levels and imperfect phase stepping make reconstruction challenging with conventional techniques. We compare four reconstruction methods on simulated and experimental MAI-SIM data. FairSIM[5] and the inverse matrix[6] methods represent implementations of original SIM reconstruction algorithm proposed by Gustafsson.[7] In both cases the pattern periods and orientations are determined by iterative cross correlation of the Fourier transforms of the data and the two techniques differ by the phase estimation approaches used. By assuming an equidistant phase separation (i.e. phases separated by $2\pi/3$), FairSIM reduces the parameter estimation process to extraction of only a single global phase offset. In contrast, the inverse matrix approach recovers the phases by analytically solving a trigonometric linear equation without the need for iterative optimization, recovering absolute phase values for each image. Frequency reassignment is performed with the same method for both FairSIM and the inverse matrix approach and the methods differ only in the phases used for this reassignment. The blind SIM approach used was a MATLAB implementation of a joint reconstruction strategy where both the reconstructed image and the illumination pattern are iteratively optimized from initial estimates.[8] The initial estimate for the sample was taken as the average of the nine raw SIM frames and an initial estimate of the pattern was calculated by element-wise division of the sample estimate by each raw frame.

The machine learning approach used to address uneven phase stepping is based on previously described work using transfer learning to train a convolutional neural network (CNN) on simulated SIM images.[9] Training data were generated by modelling the SIM image formation process on artificial "samples" taken as 512x512 pixel patches from images in the high resolution DIV2K image dataset.[10] The imaging parameters for pixel-size; excitation and emission wavelengths; fringe spacing, orientation and modulation depth; and imaging depth were randomized within the bounds of the expected imaging conditions. Random Poisson and Gaussian noise were then added to the images. The model was then trained in a supervised fashion where the original image patches were used as the ground-truth. The Adam optimizer and mean-square error were used for training and the model was trained on 5000+ samples for 300 epochs. Python code and the PyTorch[11] machine learning library were used throughout. The code used to generate the training data, train the ML-SIM models and evaluate the model on experimental data is available in the GitHub [repository](). Three models with different network architectures are available for evaluation in this repository. The first is a broad model with 96 feature maps, 10 residual blocks and 3 residual groups.[9] The second is a narrower model which can be evaluated faster consisting of 48 feature maps, 10 residual blocks and 3 residual groups. Live ML-SIM reconstructions were performed with the shallower model for best execution speed. Static ML-SIM reconstructions were performed using the broader model and reconstructions of dynamic structures were performed after acquisition using an experimental visual transformer model which has improved performance on dynamic structures at the expense of increased reconstruction times.[12]

To evaluate the performance of the techniques on data with uneven phases steps we first quantitatively measure the Resolution Scaled Pearson (RSP) coefficient and the Resolution Scaled Error (RSE) implemented through the NanoJ plugin for imageJ.[13] These provide a measure of the likelihood that a reconstruction could represent the real sample based on comparison between a predicted and real diffraction limited image. RSP values range from 0 to 1, with 1 representing a perfect correlation and ideal reconstruction. The RSE is a measure of the mean square error values between the predicted widefield and the observed widefield with a lower value showing a better reconstruction. Table S1 compares the performance of the reconstruction techniques on simulated SIM data with ideal and imperfect pattern phase stepping. The use of simulated targets allows for the effects of only phase errors to be assessed without other factors (such as background fluorescence and artefacts in the reference image) affecting the reconstruction quality metric. The code to generate the simulated images can be found on the project [GitHub]() and the images used can be found in the raw [data](). In all cases the reconstructions have performed better on data with the correct phase steps, which is expected. Notably, ML-SIM out-performs all methods except for the inverse matrix method when ideal phase shifts are used. For this reason, the inverse matrix method was used for sequential data and ML-SIM reconstruction was used on simultaneous acquisitions.

*Table S1: Reconstruction quality measured by RSP and RSE.* For all reconstruction techniques, reconstructions are better on data with correct phase stepping. In the case of FairSIM this is expected as phases cannot be calculated. For Blind SIM, this discrepancy can be attributed to the initial estimate of the sample containing residual fringe patterns. For the inverse matrix and ML-SIM methods, reduced performance is likely a result of reduced information being available on regions of the sample unevenly illuminated or residual striping in the widefield image.

| Method | Bad phases | | Good phases | |
|---|---|---|---|---|
| | RSP | RSE | RSP | RSE |
| ML-SIM | 0.991 | 1497 | 0.994 | 1134 |
| Blind SIM | 0.978 | 2280 | 0.983 | 1960 |
| FairSIM | 0.975 | 2419 | 0.992 | 1362 |
| Inverse Matrix | 0.980 | 2196 | 0.995 | 1085 |

Live ML-SIM reconstruction was achieved by using the Pycromanager[14] and NI-DAQmx libraries in Python. A multi-threaded approach was implemented whereby image acquisition, reconstruction and display were performed in parallel. The GUI to perform real-time reconstructions is shown in Figure S8 Full details on the implementation can be found in the documentation available for the MAI-SIM software and a video of GUI operation is found in the data repository.

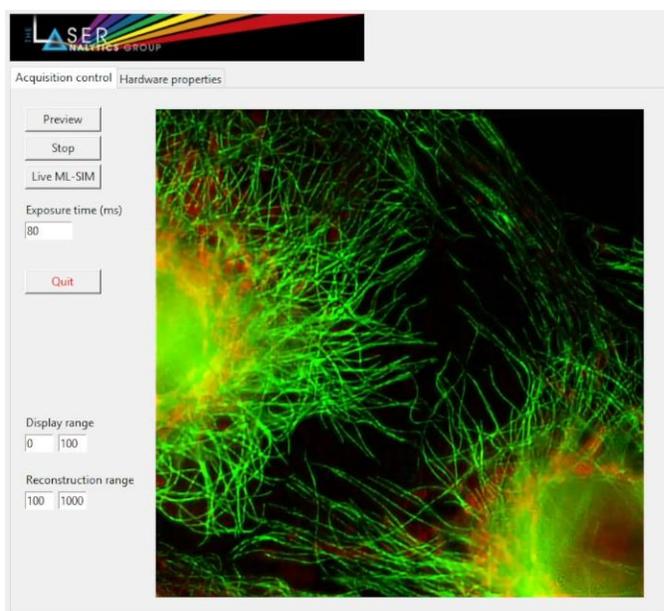

*Figure: S8 GUI for live ML-SIM reconstruction.* Image shows a 2-color reconstruction of fixed COS7 cells over a 44x44 μm field of view. Immunostained AlexaFluor647-Tubulin and GPF-tagged mitochondria are shown in green and red respectively. Video of operation can be found in the supplementary data.

# Supplementary Note 3: Sample preparation

**Bead preparation:** Bead samples were prepared according to manufacturer protocols. Briefly 200 nm and 100 nm bead monolayers (ThermoFisher, TetraSpeck) were formed by air-drying onto a 12 mm coverglass 5 µl of bead suspensions at 2.3 x $10^7$ and 1.8 x $10^8$ particles/mL respectively. The samples were then mounted in 15 µl DI water and secured to glass slides with spacers before being sealed with nail polish.

**Fixed cell imaging:** COS-7 cells (from monkey kidney tissue) were plated onto 13 mm round coverslips in a 4 well plate at 20,000 cells per well and cultured under standard conditions (37°C, 5% CO2) in minimum essential medium (Sigma Aldrich) supplemented with 10% foetal bovine serum (Gibco) and 2 mM L-lutamine (GlutaMAX, Gibco). The next day, cells were transfected using CellLight Mitochondria-GFP (Invitrogen), dispersing 8 µL of the transfecting reagent in the cell culture medium and incubating the cells for 24 hours under standard conditions. Cells were fixed by incubation with 4% methanol-free formaldehyde and 0.1% glutaraldehyde in cacodylate buffer (pH 7.4) for 15 minutes at room temperature, washed three times with PBS and then permeabilized by incubation with a 0.2% solution of saponin in PBS for 15 minutes. Unspecific binding was blocked by incubating with 10% goat serum and 100 mM glycine in PBS and 0.2% saponin for 30 minutes at room temperature. Without washing, the samples were incubated with the primary antibodies (mouse anti-beta-tubulin: ab131205, rabbit anti-calnexin: ab22595) diluted 1:200 in PBS containing 2% BSA (bovine serum albumin) and 0.2% saponin overnight at 4°C. After three washes in PBS, the samples were incubated with the secondary antibodies (goat anti-mouse conjugated to AlexaFluor568, goat anti-rabbit conjugated to AlexaFluor647) diluted 1:400 in PBS containing 2% BSA and 0.2% saponin for 1 Hour at room temperature in the dark. Samples were then washed 3 times with PBS. Finally, the coverslips were mounted on glass slides using a Mowiol-based mounting medium.

**Live cell imaging of ER-lysosome-mitochondria dynamics:** COS-7 cells were plated into 8 well plates (Ibidi) at 5,000 cells per well and cultured under standard conditions (37°C, 5% CO2) in Dulbecco's Modified Eagle's Media (DMEM, ThermoFisher) supplemented with 10% foetal bovine serum (Gibco) and 2 mM L-glutamine (GlutaMAX, Gibco). 24 hours before imaging, cells were transfected with the plasmid construct mApple-Sec61β-C1 (Addgene plasmid #90993) using Lipofectamine 2000 (ThermoFisher) according to the manufacturer's protocol. 20 hours prior to imaging cells were stained with SiR-Lysosome (Cytoskeleton Inc.) and Verapamil (Cytoskeleton Inc.) at 1 µM and 10 µM respectively. 30 minutes prior to imaging cells were stained with MitoTracker Green FM (Thermofisher) at 250 nM and washed immediately before imaging with DMEM. Imaging was performed with a stage-top incubator (OKOLab) at 37°C and 5% CO2.

# Supplementary Note 4: Extension to 3D-SIM

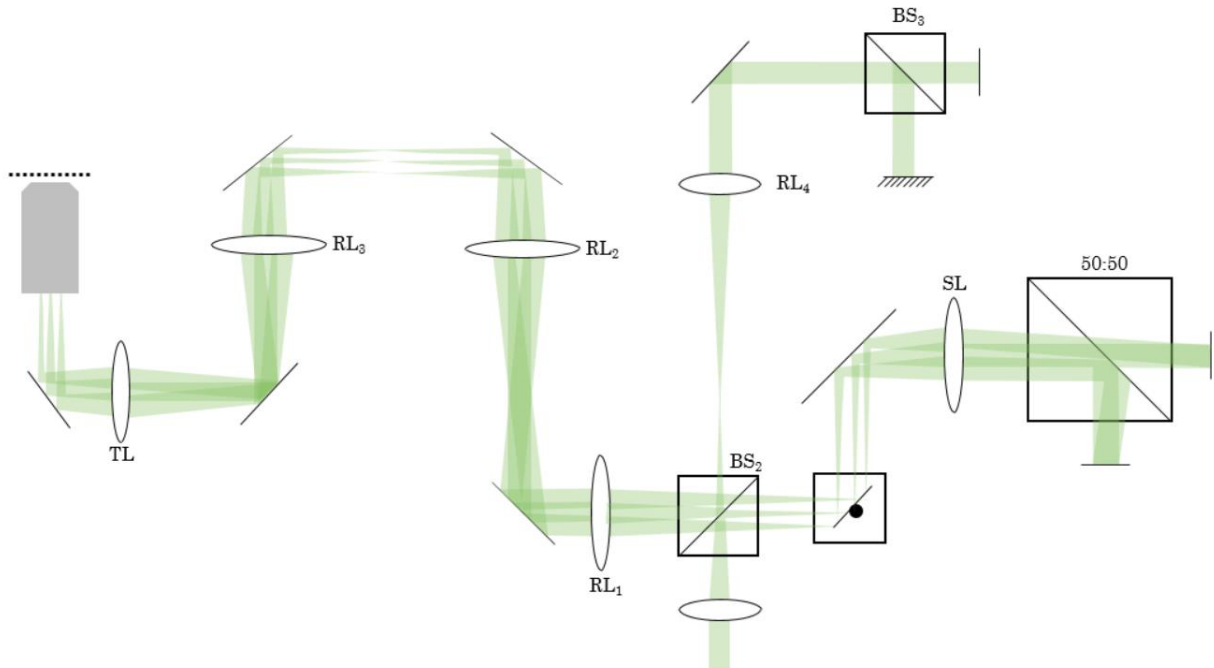

*Figure: S9 Proposed optical setup for 3D MAI-SIM. By replacing the D-mirror with a second beamsplitter cube (**BS₂**) and additional arm can be introduced to the interferometer to allow for 3-beam interference in the sample. A third beamsplitter (**BS₃**) preserves the path length of the beams and ensure equal beam intensity.*